\documentclass[10pt,letterpaper]{article}
\usepackage[top=0.85in,left=2.75in,footskip=0.75in]{geometry}
\usepackage{amsmath,amssymb}
\usepackage{changepage}
\usepackage[utf8x]{inputenc}
\usepackage{textcomp,marvosym}
\usepackage{cite}
\usepackage{nameref,hyperref}
\usepackage[right]{lineno}
\usepackage{microtype}
\DisableLigatures[f]{encoding = *, family = * }
\usepackage[table]{xcolor}
\usepackage{array}
\newcolumntype{+}{!{\vrule width 2pt}}
\newlength\savedwidth

\raggedright
\usepackage[aboveskip=1pt,labelfont=bf,labelsep=period,justification=raggedright,singlelinecheck=off]{caption}

\makeatletter
\renewcommand{\@biblabel}[1]{\quad#1.}
\makeatother

\usepackage{lastpage,fancyhdr,graphicx}
\usepackage{epstopdf}
\fancyheadoffset[L]{2.25in}
\fancyfootoffset[L]{2.25in}
\begin{document}
\vspace*{0.2in}

\begin{flushleft}
{\Large
\textbf\newline{ANGUS: Real-time manipulation of vocal roughness for emotional speech transformations} 
}
\newline
\\
Marco Liuni\textsuperscript{1,2},
Luc Ardaillon\textsuperscript{1},
Louise Bonal\textsuperscript{3},
Lou Seropian\textsuperscript{4},
Jean-Julien Aucouturier\textsuperscript{1}*,
\\
\bigskip
\textbf{1} Science \& Technology of Music and Sound (STMS), IRCAM/CNRS/Sorbonne Université, Paris, France.
\\
\textbf{2} Alta Voce SAS, Paris, France.
\\
\textbf{3} Formerly at HES-So Valais, Switzerland.
\\
\textbf{4} Lyon Neuroscience Center, Lyon, France. 
\\
\bigskip

* aucouturier@gmail.com

\end{flushleft}
\section*{Abstract}
Vocal arousal, the non-linear acoustic features taken on by human and animal vocalizations when highly aroused, has an important communicative function because it signals aversive states such as fear, pain or distress. In this work, we present a computationally-efficient, real-time voice transformation algorithm, ANGUS, which uses amplitude modulation and time-domain filtering to simulate roughness, an important component of vocal arousal, in arbitrary voice recordings. In a series of 4 studies, we show that ANGUS allows parametric control over the spectral features of roughness like the presence of sub-harmonics and noise; that ANGUS increases the emotional negativity perceived by listeners, to a comparable level as a non-real-time analysis/resynthesis algorithm from the state-of-the-art;  that listeners cannot distinguish transformed and non-transformed sounds above chance level; and that ANGUS has a similar emotional effect on animal vocalizations and musical instrument sounds than on human vocalizations. A real-time implementation of ANGUS is made available as open-source software, for use in experimental emotion reseach and affective computing.



\section*{Introduction} 
\label{sec:introduction}


When alarmed, threatened or highly aroused, the vocalizations of humans and many non-human animals take on a number of non-linear acoustic features, such as subharmonics and broadband noise, which give them a rough and noisy sound quality \cite{FIT02}. Vocal arousal in screams, cries, grunts or moans has an important communicative function in the human expressive repertoire, because it signals aversive states such as fear, pain or distress \cite{RAINE18} and more generally modulates the perceived valence, arousal and dominance of emotional vocalizations \cite{ANIK19-2}. The perception of vocal arousal triggers fast and stereotypical responses in listeners \cite{ASUT17} and evokes activity in areas linked to the brain's threat response system \cite{ARNA15}. Beyond language, vocal arousal and non-linearities are also prominently featured in popular music, e.g. in the growling vocal style of jazz or popular music singers such as Louis Armstrong, Tom Waits or Kurt Cobain \cite{SAKA04} and, perhaps most remarkably, in extreme music such as metal \cite{OLL19}. 

From a computational point of view, the analysis and synthesis of vocal arousal is important for all experimental systems concerned with negative emotional states, such as anger (e.g., detecting aggression in speech - \cite{SAH16}), fear (e.g., using screams to induce anxiety experimentally - \cite{BEAU19}) or pain (e.g., automated medical assessment - \cite{HAM12}). For voice analysis, the measurement of the specific non-linear features of vocal arousal is often formulated in terms of jitter and shimmer (cycle-to-cycle variations in the period and amplitude of glottal pulses, respectively) and harmonic-to-noise ratio (HNR) \cite{BACH95}. However, the control of nonlinear vocal behaviour in synthesized or existing recordings is less well-established. Classical glottal models, which represent individual glottal pulses as a parametric function \cite{FANT85, VELD98}, or additive sinusoidal models, which generate a separate sine wave for each harmonic \cite{ANIK19,SUEUR08}, can be modified to generate variability in the glottal shape or periodicity. For instance, to create additional subharmonics, one can use a second glottal model at a pseudo-harmonic frequency to modulate the main series of pulses \cite{ALON15} or add additional components at an integer ratio of the fundamental frequency \cite{ANIK19}; to add jitter or shimmer, one can add stochastic variations to the phase or amplitude of sinusoidal components \cite{FRAJ12}. However, these synthesis techniques do not make it easy to modify pre-recorded vocal signals (e.g. dynamically modulate the arousal of video game character's scream, add growl on a singer's voice) or voices synthesized with methods such as concatenative text-to-speech \cite{HUNT96} or sample-based deep network methods \cite{OORD16}. Adding vocal non-linearities on such signals can be done with analysis-resynthesis techniques that, first, estimate the recording's series of glottal pulses \cite{DEGOT13} and, then, resynthesize the vocal signal from a manipulated series of pulses with artificially-varied amplitude and period \cite{VERM05,RUIN08,BOHM08,ARD17}. However, these techniques rely on an explicit model of pulse variability, which is typically learned from one or several target examples of naturally rough voices \cite{BON13}, and it is unclear how such predetermined patterns should be selected for unknown voices. Moreover, because of the computational complexity of the initial stage of glottal source estimation, these techniques cannot operate in real-time, i.e. dynamically manipulate vocal arousal in an incoming vocal stream, thus ruling out most applications in continuous, dyadic interactions \cite{AUC20}. 

Here, we present a new acoustic transformation algorithm, called ANGUS\footnote{\url{http://forumnet.ircam.fr/product/angus}}, which aims to simulate the rough timbral quality of vocal arousal without attempting to directly control jitter and shimmer at the glottal source level. The algorithm, which uses amplitude modulation and time-domain filtering to add sub-harmonics to the original signal, is computationally extremely efficient and operates in real-time. In the following, we compare the algorithm to a state-of-the-art non-real-time alternative method based on pulse resynthesis \cite[Section 6.3.2]{ARD17} and evaluate (1) its ability to parametrically (albeit indirectly) modulate a vocal signal's jitter and shimmer, (2) its impact on listener's judgements of the emotional negativity, (3) on judgements of transformation naturalness and (4) its ability to transform both vocal and non-vocal (musical, environmental) sounds. 

\section*{Algorithms}
\subsection*{ANGUS}
Our proposed approach to model vocal arousal uses amplitude modulation and time-domain filtering
to efficiently add sub-harmonics in the original signal. Amplitude modulation consists in multiplying in the time-domain a carrier signal with a modulating signal that has a lower frequency and an amplitude varying in the range $[0,1]$, centered around 1. 
Let $x_c(t)=A_c cos(\omega_ct)$ be the carrier signal, with angular frequency $\omega_c$ and amplitude $A_c$, and $x_m(t)=1 + hcos(\omega_mt)$ be the modulating signal with angular frequency $\omega_m$ and modulation depth $h\in [0,1]$. We then have: 
\begin{equation}
\label{eqn:1}
\begin{split}
y(t) & = x_m(t)x_c(t)\\
& = (1 + h cos(\omega_mt)) A_c cos(\omega_ct)\\
& = A_ccos(\omega_ct) + A_chcos(\omega_mt)cos(\omega_ct)\\
& = x_c(t) + \frac{A_ch}{2}cos((\omega_c + \omega_m)t) + \frac{A_ch}{2} cos((\omega_c - \omega_m)t)\\
&= x_c(t) + y^+(t) + y^-(t)
\end{split}
\end{equation}

\begin{figure*}[ht]
\centering
 \includegraphics[width=\linewidth]{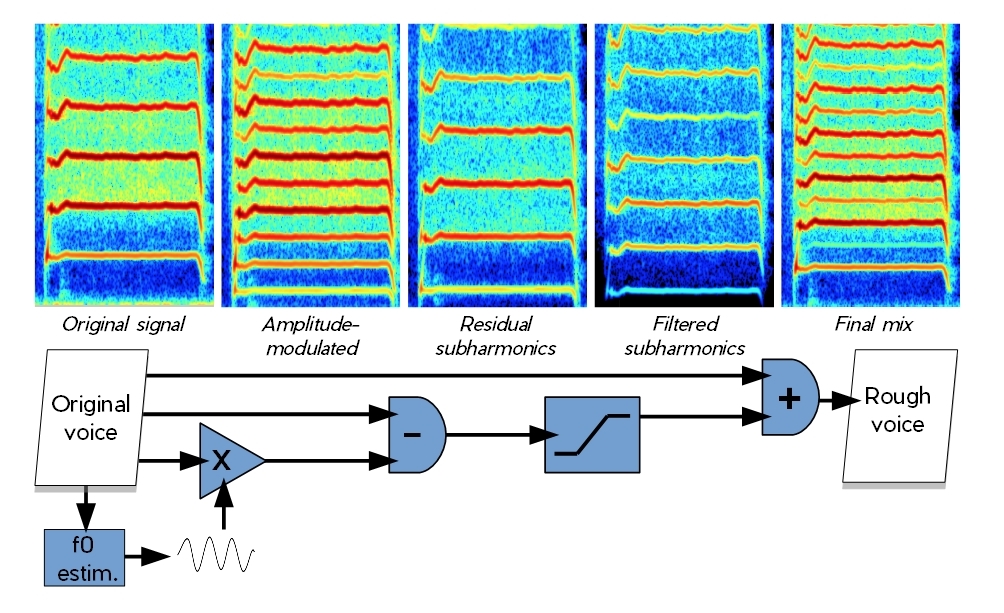}
 \caption{{\bf The ANGUS algorithm.} ANGUS simulates vocal arousal using amplitude modulation and time-domain filtering to efficiently add sub-harmonics in the original signal.Top: Effect of each processing step of the algorithm on the spectrogram of one male vocalization.Bottom: Algorithm workflow.}
 \label{algo}
\end{figure*}

The resulting signal contains the original signal $x_c(t)$, and two new sinusoids: $y^+(t)$ and $y^-(t)$ at frequency $\omega_c+\omega_m$ and $\omega_c-\omega_m$. In case $x_c(t)$ is a voice signal at fundamental frequency $f_0$, approximated by a sum of N harmonics: $x_c(t)=\sum_{i=1}^{N}A_icos(i\omega_0t)$, where $\omega_0=2\pi f_0$, the modulation of this signal by $x_m(t)$ is the sum of each harmonic modulated individually:
\begin{equation}
\label{eqn:2}
\begin{split}
y(t) & = \sum_{i=1}^{N} x_m(t)A_icos(i\omega_0t)\\
& = x_c(t) + \sum_{i=1}^{N} y_i^+(t) + y_i^-(t)\\
\end{split}
\end{equation}
where 
\begin{equation}
\label{eqn:3}
\begin{split}
y_i^+(t) &= \frac{A_ih}{2}cos((i\omega_0 + \omega_m)t)\\
y_i^-(t) &= \frac{A_ih}{2}cos((i\omega_0 - \omega_m)t)
\end{split}
\end{equation}
By choosing appropriate values for $\omega_m$, it is thus possible to generate sub-harmonics $y_i^+$ and $y_i^-$ between each of the original signal's partials, at frequencies $\omega_0\pm\omega_m$. In particular, setting
$\omega_m = \omega_0/k$ generates pairs of sub-harmonics around each harmonic at $i\omega_0\pm\omega_0/k$. This is the approach we use here. In this work, we will evaluate the impact of adding sub-harmonics at $\omega_0/k$ with $k\in\{2,3,4,5\}$ . 

In addition, because amplitude modulation applies the same amplitude factor $h$ to all created sub-harmonics (Eq.\ref{eqn:2}), a provision is added to avoid unrealistically high amplitude for the lowest of these sub-harmonics, by high-pass filtering the sub-harmonic residual. As the original signal $x_c(t)$ is fully preserved in the modulated
signal (Eq.\ref{eqn:2}), the sub-harmonics can be
isolated by subtracting this original signal from the
modulated one: $y_{sub}(t) = y(t) - x_c(t)$. Once the subharmonics are isolated, they are filtered with a second order bi-quad high-pass filter, with cutoff frequency $f_{cut} = 4f_0$, before being added back to the original signal by simple summation. We thus obtain the final rough voice signal as: $y_{rough}(t) = x_c(t) + \alpha HP(y_{sub}(t))$, where $HP(y_{sub}(t))$ denotes the high-pass filtered sub-harmonics and $\alpha > 0$ is a mixing factor. In this work, we will evaluate the impact of $\alpha$ for the values  $\alpha\in\{.25,.5,.75,1\}$ (modulation depth fixed to $h=1$). 

A real-time implementation of this algorithm, as an open-source patch (MIT License) based on the closed-source software Max (Cycling '74), as well as a series of sound examples, are available at \url{http://forumnet.ircam.fr/product/angus} \cite{LIUNI18}. The only computations required for the transformation are one multiplication for amplitude modulation, one subtraction to isolate sub-harmonics, a few  multiplications and additions for filtering (depending on the order of the filter), and one addition and one multiplication for the final mixing step. This thus makes this approach especially suitable for real-time. 

\subsection*{Control algorithm}
\label{control}

In this work, we compare the ANGUS algorithm to a more sophisticated, but non-real-time state-of-art method based on pulse analysis and resynthesis \cite[Section 6.3.2]{ARD17}. Contrary to ANGUS, this control algorithm requires the availability of original rough recordings from which a pattern of jitter and shimmer can be analysed. The algorithm then resynthesizes the original recording with varying levels of jitter and shimmer. 

The control algorithm (thereafter, CONTROL) uses the \textit{PaN} parametric speech synthesis engine \cite[Section~3.5.2]{ARD17} to precisely control the positions and amplitudes of synthesized pulses. In more details, CONTROL first uses peak-picking to identify each individual pitch cycles, and computes the time series of their inter-peak-intervals (the  variability  of  which  is  related  to  jitter) and inter-peak-amplitude-differences (related to shimmer). Then, it creates a new series of synthetic glottal pulses using the \textit{PaN} synthesis engine, by interpolating at a ratio $\alpha_c$ between the original jitter and shimmer time-series ($\alpha_c$=1.0) and a strictly periodic and iso-amplitude series set at the local average (slowly varying) period and amplitude of the original series ($\alpha_c$=0). Finally, the vocal signal is obtained by filtering the pulses with the corresponding vocal tract filter estimated on the original recording using the true-envelope algorithm \cite{Roebel2005}, and by adding the high-frequency noise component extracted from the original recording (above the highest prominent harmonic). In this work, we use mixing factors $\alpha_c \in \{0,0.25,0.5,0.75,1.0\}$, where $\alpha_c$=1.0 corresponding to levels of jitter and shimmer that are identical to that of the originally rough signals.  

Note that, in the following, stimuli compared between the ANGUS and CONTROL procedures were not strictly equivalent: while we use ANGUS to add subharmonics to clear voice, thus increasing roughness, we use the CONTROL algorithm to decrease the roughness of originally rough voice (see Study 1-4-{\it Procedure}). In other words, both algorithms were not applied to identical stimuli, but to pairs of clear/rough recordings by the same speaker. The purpose of this procedure is therefore not to evaluate ANGUS as a computational alternative to the CONTROL algorithm, but to provide a set of reference stimuli in which jitter and shimmer are parametrically controlled and to evaluate how well ANGUS is able to simulate the higher-level perceptual-cognitive aspects of these stimuli, in terms of scalability (Study 1), perceived valence (Study 2), and perceived naturalness (Study 3), in addition to its transferability to non-vocal stimuli (Study 4), which is not possible with CONTROL.

\section*{Study 1: Effect on jitter and shimmer}

Although ANGUS does not directly model pulse frequency and amplitude, the signal manipulations used here are indirectly related to shimmer and jitter. First, amplitude modulations with modulation frequency $\omega_m$ proportional to $\omega_0$ are expected to create periodic patterns of shimmer. Second, while sub-harmonics are not expected to create variations of pulse periodicity (although, conversely, increased jitter would create sub-harmonics \cite{GIO99}), the modifications imposed to the waveform by ANGUS may also affect the estimation of pulse positions used to measure jitter in typical implementations (such as Praat \cite{BOER10}). Voice processed with ANGUS are therefore expected to have increased levels for both measures. To quantify this effect, we evaluate here the effect of the $k$ and $\alpha$ parameters of ANGUS on vocal jitter and shimmer, and compare with that of the $\alpha_c$ parameter of the CONTROL algorithm, which is designed to directly and linearly modulate these two measures.  

\subsection*{Stimuli}
Stimuli for the experiment consisted of 24 short, 1-second recordings of human vocalizations (12 neutral, 12 rough). Vocalizations were recorded by one female (F1) and two male (M1,M2) actors instructed to shout/sing phonemes [a] (for F1,M1,M2) and [i] (for F1) at three different pitches (in the range [450,480], [570,600] and [520,570] Hz for F1; [200,215], [250,270] and [315,340] Hz for M1 and M2), with a clear, loud voice (neutral stimuli), and to scream the same material with an angry, excited voice (rough stimuli). The resulting 24 stimuli (F1a,F1i,M1a,M2a $\times$ 3 pitches $\times$ neutral/rough) were then manually segmented to their middle 600ms sustained section, cutting onset and decay, and normalized in loudness. Original neutral files (12 stimuli) were processed with ANGUS at 4 levels of $\alpha\in\{.25,.5,.75,1\}$, and a single modulating signal at different subharmonic ratio $\frac{\omega_0}{k}, k\in\{2,3,4,5\}$, resulting in 192 transformed files. Original rough stimuli (9 stimuli; for technical issues, M2a files were not processed) were transformed with CONTROL, at 4 levels of mixing factor $\alpha_c\in\{.25,.5,.75,1.0\}$, resulting in 36 stimuli. 

\subsection*{Procedure} 
Local jitter (average absolute difference between consecutive periods, divided by the average period) and shimmer (average absolute difference between the amplitudes of consecutive periods, divided by the average amplitude) were measured from recordings transformed with ANGUS and CONTROL using the Praat software \cite{BOER10}. Statistical effects of the algorithm parameters on both measures were inferred from the sample of original recordings (N=12) using a rmANOVA, with $k$,$\alpha$ (ANGUS) and $\alpha_c$ (CONTROL) as within-item factors; statistics reported below with {\it F} and {\it p} values. 

\subsection*{Results}
As expected, CONTROL had a significant, linear impact on the jitter (F(1,8)=8.69, p=.018) and shimmer (F(1,8)=23.51, p=.001) of the transformed sounds, which culminated at $\alpha_c=1.0$, with a mean jitter value of 1.4\% (above the 1.040\% threshold for pathological speech, as expected for the scream register \cite{BOER10}) and mean shimmer value of 15\% (above the 3.8\% threshold for pathological speech). 

Although it did not directly model these measures, ANGUS also had a significant effect on measured jitter, which increased both with $\alpha$ (F(1,11)=6.63, p=.025) and $k$ (F(1,11)=5.45, p=.039), and shimmer ($\alpha$: F(1,11)=7.53, p=.019; $k$: F(1,11)= 19.80, p=.0009). Maximum jitter values (ca. 1) were reached for $k\geq4$ and $\alpha=1.0$ while maximum shimmer (ca. 10) was reached for $k=3$; both were comparable, but below the levels reached by CONTROL at $\alpha_c$=1.0, i.e. the ANGUS effect at $\alpha = 1$ resulted in transformations that had less measured jitter and shimmer than the originally rough vocalizations of our corpus.

\begin{figure*}[ht]
\centering
 \includegraphics[width=\linewidth]{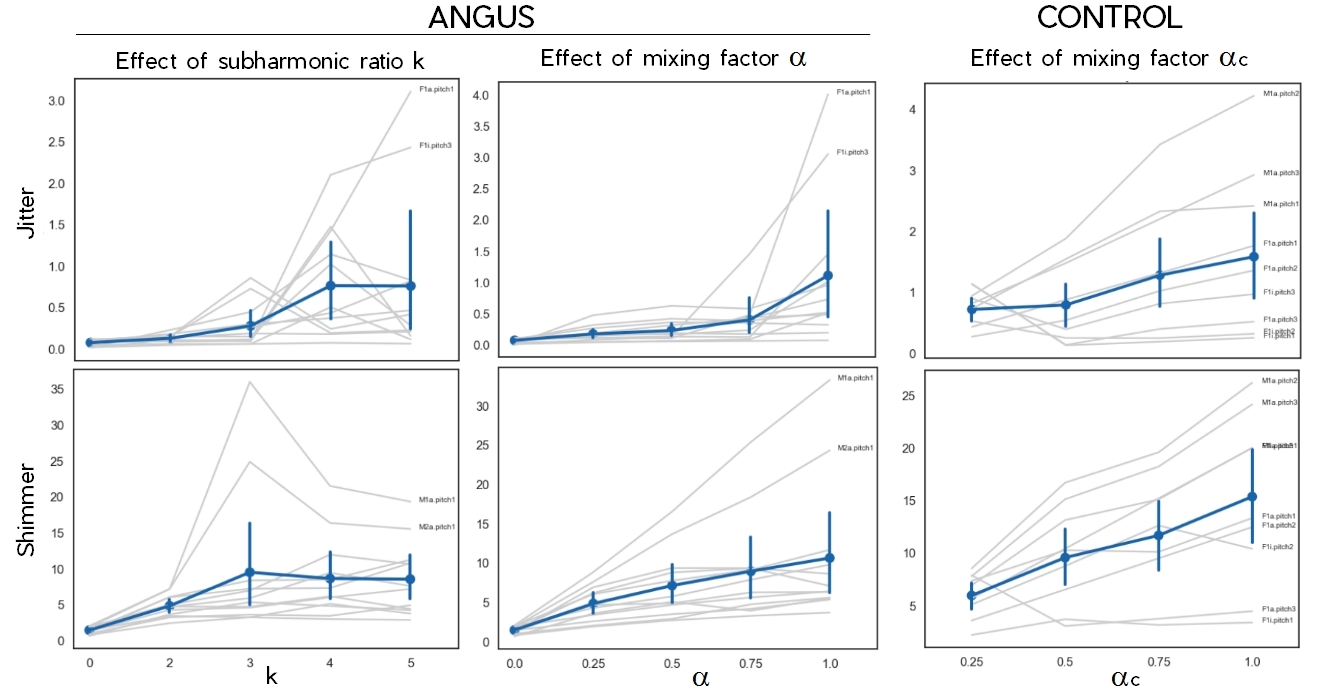}
 \caption{{\bf Effect of ANGUS and CONTROL on the measured jitter and shimmer of transformed extracts.} Left: ANGUS $k$ had a significant effect on jitter (top) and shimmer (bottom), with maximum values reached for $k$=4 (jitter) and $k$=3 (shimmer). Middle: ANGUS $\alpha$ had a significant impact on jitter and shimmer, as did CONTROL $\alpha_c$ (Right), although the ANGUS effect at maximum amplitude resulted in transformations that were less intense than the originally rough vocalizations of our corpus. In all graphs, light grey lines correspond to individual recordings, and solid blue lines to the average over all recordings. Error bars: 95\% CI on the mean.}
 \label{acoustics}
\end{figure*}

\section*{Study 2: Effect on perceived negativity}

\subsection*{Participants}
N=21 young French speaker (mean age M=22.9, SD=2.6; 12 female, 9 male) with normal hearing took part in the experiment. Participants signed an informed consent form, and were paid for their participation. All experiments in the following were approved by the INSEAD-Sorbonne Universit\'e Internal Review Board (IRB). 

\subsection*{Stimuli}
Stimuli were the same as in Study 1, and consisted of 24 original neutral recordings (12 neutral, 12 rough), 192 transformed stimuli obtained from the original neutral recordings by manipulation with ANGUS (at various levels of $\alpha$ and $k$), and 36 transformed stimuli obtained from the originally rough recordings by manipulation with CONTROL (at various levels of $\alpha_c$).

\subsection*{Procedure}
Study 2,3 and 4 were conducted in a single experimental session, in the order 2-4-3. In Study 2, participants were presented with all 252 stimuli in randomized order. In the current study (as well as Study 4), instructions given to participants implied that all stimuli were genuine (non-transformed) recordings of human vocalizations. In each trial, participants had to evaluate the emotion communicated by the speaker, using a unipolar continuous scale anchored by 1, corresponding to a sung note with no emotion and 10, corresponding to a highly negative and highly aroused expression. For each participant, ratings of negativity were averaged over all extracts that were transformed with the same parameter settings, and statistical effects of the algorithm's parameters were inferred from the sample of participants (N=22) using a rmANOVA, with pitch, $k$,$\alpha$ (ANGUS) and $\alpha_c$ (CONTROL) as within-item factors. 

\subsection*{Results}

\begin{figure*}[ht]
\centering
 \includegraphics[width=\linewidth]{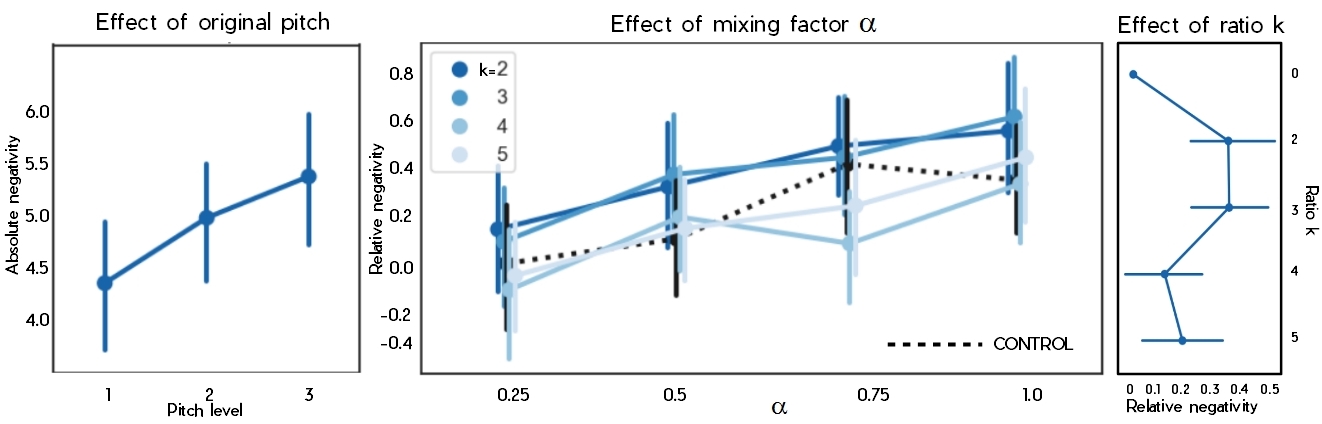}
 \caption{{\bf Effect of vocalization pitch and ANGUS parameters on perceived negativity}. Left: High-pitch vocalizations were judged more negative than low-pitch vocalizations. Middle: ANGUS $\alpha$ (solid lines) had a linearly increasing effect on negativity, comparable in strength to that of the CONTROL algorithm (dashed line). Right: Subharmonic ratios $k$=2 and 3 had the strongest effect of negativity. Error bars: 95\% CI on the mean.}
 \label{neg}
\end{figure*}

Both the $k$ and $\alpha$ parameters of ANGUS had strong effects on the perceived negativity of the transformed extracts (Figure \ref{neg}-Middle \& Right). Subharmonic ratio $k$ (F(3,60)=12.38, p=2e-06) had a stronger effect for $k\in\{2,3\}$ (M=5.0) than $k\in\{4,5\}$ (M=4.8). At $k$=3, the increase of negativity over that of neutral recordings was $\Delta$=+0.38 point of scale (95\%-CI: [0.31,0.45]). Mixing factor $\alpha$ (F(3,60)=33.89, p=6e-13) had a linearly increasing effect on negativity (Figure \ref{neg}-Middle), with $\alpha$=1.0 resulting in a $\Delta$=+0.49 [0.41,0.57] point increase over the negativity of corresponding neutral extracts.

For comparison, we examined the effect of the $\alpha_c$ mixing factor of the CONTROL algorithm. There was a significant effect on perceived negativity (F(3,60)=8.08, p=.0001), which was comparable in amplitude to that of ANGUS $\alpha$ (Figure \ref{neg}-Middle). At $\alpha_c$=1.0, the mean increase of negativity over that of neutral extracts was $\Delta$=0.42 [0.28,0.56].

Independently from algorithm parameters, there was a strong effect of vocalization pitch on perceived negativity (F(2,40)=40.42, p=2e-10). High-pitch vocalizations (M=5.3) were judged more negative and aroused than low-pitch vocalizations (M=4.4; $\Delta$ = +0.99 [+0.83,1.14]; Figure \ref{neg}-Left).

\section*{Study 3: Perceived naturalness}

\subsection*{Participants}
Participants were the same as in Study 2 (N=21). Study 2,3 and 4 were conducted in a single experimental session, in the order 2-4-3, i.e. Study 3 was the last task. 

\subsection*{Stimuli}
Stimuli consisted of 36 non-manipulated extracts and 36 manipulated extracts, incl. 12 ANGUS recordings and 24 CONTROL recordings. Stimuli were obtained from the same original recordings as Study 1, in the following manner: the 36 non-manipulated stimuli were obtained from the 12 original rough recordings (F1a,F1i,M1a,M2a $\times$ 3 pitches), each segmented into 3 different 600ms sections at various positions into the file, i.e. they were physically different extracts of the same utterances. The 12 ANGUS stimuli were obtained from the 12 original neutral recordings by transformation at the fixed setting $k$=3 and $\alpha$=.75. The 24 CONTROL stimuli were obtained from the 12 original rough recordings (F1a,F1i,M1a,M2a $\times$ 3 pitches) by transformation at $\alpha_c$=.75. Twelve of these stimuli were obtained using the jitter and shimmer profile of the original rough recording of the same speaker (e.g. F1a $\leftarrow$ F1a, as described in Section \ref{control}). In addition, twelve other CONTROL stimuli were obtained by cross-synthesis, using the jitter and shimmer profile of another rough original recording in the same dataset (F1a$\leftarrow$F1i, $\times$ 3 pitches; F1a$\leftarrow$M2a, $\times$ 3 pitches; F1i$\leftarrow$F1a, $\times$ 3 pitches; M2a$\leftarrow$M1a, $\times$ 3 pitches). 

\subsection*{Procedure}
Upon completion of Study 2 and 4, and before the current study, it was revealed to participants that some of the sounds had in fact been artificially manipulated to sound negative and aroused. Participants were then presented with all 72 stimuli in randomized order and explained that about half of these stimuli were manipulated to sound negative and aroused, and half were original recordings of speakers embodying these emotions. At each trial, participants had to indicate whether they thought the sound was original or transformed. We evaluated the algorithms' naturalness by computing each participant's hit rate and sensitivity index {\it d'} (computed with the log-linear correction \cite{HAU95}) corresponding to the successful detection of the 36 manipulated trials (chance level= 0.5).  

\subsection*{Results}
The detection hit rate for ANGUS was statistically above chance level (H=0.58, [0.52,0.64]) but, because of a large number of false alarms (FA=0.57, i.e. many of the original samples were believed to be manipulated),  {\it d'} was not statistically different from zero ({\it d'}=-0.005, [-0.17,0.17]). This indicates that participants were not better than chance at discriminating sounds transformed with ANGUS from genuine rough sounds. 

The detection hit rate for the CONTROL algorithm was not statistically different from chance (H=0.48, [0.41,0.54]), and {\it d'} was even significantly negative ({\it d'}=-0.31, [-0.48,-0.13]), indicating that participants were consistent in finding original sounds \emph{less} natural than sounds transformed with CONTROL. Correspondingly, there was a significant effect of algorithm ANGUS or CONTROL on participant {\it d'} (F(1,20)=6.87, p=.016), indicating that CONTROL stimuli sounded more natural to participants than ANGUS stimuli, but only because participants paradoxically found that CONTROL sounds were more natural than original sounds.  

\section*{Study 4: Effect on non-vocal stimuli}

One major difference between the transformation approach of ANGUS and the analysis/resynthesis approach of the CONTROl algorithm  used here, as well as other approaches for jitter modeling in the literature \cite{DEGOT13, VERM05, RUIN08, BOHM08}, is that ANGUS makes no other assumption on the input signal as its having identifiable $f_0$. Thus ANGUS can be applied to non-human vocal sounds, such as animal vocalizations, as well as non-vocal sounds such as musical instruments or alarms. There is a vast biological literature suggesting that roughness is a signal of threat and negative salience throughout the human and animal communicative repertoire \cite{FITCH02,BLUM09,ANIK18}. In addition, these signals, shaped by biological evolution, have also been selected for in cultural artefacts, such as musical sounds and sound alarms, which aim at the same effect \cite{BRYAN13,OLL19}. In \cite{BLUM12}, musical sequences transformed with distortion were evaluated as more negative and aroused than non-transformed sequences; in \cite{ARNA15}, so were musical instrument samples and alarm sounds when manipulated with 40-60Hz temporal modulations. Here, we test the capacity of a single generative model, ANGUS, to increase the perceived negativity of a variety of non-vocal sounds. 

\subsection*{Participants}
Participants were the same as in Study 2 (N=21).

\subsection*{Stimuli}
Stimuli consisted in 4 groups of 36 recordings, each composed of 12 matched manipulated and non-manipulated sounds and an additional 12 non-manipulated sounds with no manipulated equivalents. ``Scream'' stimuli included the 24 stimuli of Study 3 (F1a,F1i,M1a,M2a $\times$ 3 pitches $\times$ neutral/rough), as well as 12 non-related screams recorded in similar conditions by 2 male and 2 female additional speakers, on varied phonemes ([o]:5, [u]:5, [a]:1, [i]:1) at non-controlled pitches. ``Music'' stimuli were extracted from the McGill University Master Samples sound library \cite{MUMS89}, and included single note recordings of three wind (bugle, clarinet, trombone) and one string (violin) instrument, each performed at three different pitches. Twelve of these recordings were matched with ANGUS transformations (24 sounds: 4 instruments $\times$ 3 pitches $\times$ neutral/rough), and an additional twelve were left non-transformed. ``Animal'' stimuli were extracted from the Freesound public sound-effect database (\url{http://www.freesound.org}) and included 24 matched transformed and non-transformed recordings of animal vocalizations (cat:2, cow:3, dog:2, goat:2, sheep:3), as well as 12 unmatched additional recordings (songbird:2, cat:2, cow:1, dog:1, goat:1, rooster:2, seagull:2, swan:1). ``Object'' stimuli were extracted from the Freesound database and included 24 matched transformed and non-transformed recordings of alarms and object sounds (phone ring:1, squeaky door:2, siren:4, clock alarm:2, machine:1, spring:2), as well as 12 additional unmatched sounds (siren:4, car horn:2, spring:1, phone ring:1, clock alarm:2, machine:2). For all sounds, as in Study 3, ANGUS stimuli were obtained from the original neutral recordings by transformation at the fixed setting $k$=3 and $\alpha$=.75. 

\subsection*{Procedure}
The procedure was the same as Study 2. Participants listened to the 144 sounds, blocked by sound category (order: screams, animals, objects, music), in randomized order within each block. Unmatched sounds were included in each block (see above) to avoid attracting participants' attention to a too systematic matching of transformed and non-transformed stimuli. In each trial, participants had to evaluate the emotion communicated by the extract, using a unipolar continuous scale anchored by 1, corresponding to no emotion, and 10, corresponding to a highly negative and highly aroused emotion (we used a single construct for both emotional valence and arousal, as previous research has shown that these were identically affected by vocal roughness \cite{ARNA15,LIU20}) . Instructions differed for each type of sound as follows: for screams, participants were asked to evaluate the emotion of the speaker (as in Study 2); for animals, participants were asked to evaluate the physiological state of the animal, from calm/resting to excited/fearful/aggressive; for objects, participants were asked to evaluate to what type of situation or object usage a sound designer could use the sounds, from neutral informative situations with no urgency or danger, to highly urgent or dangerous situations that require immediate attention; for musical instruments, participants were asked to evaluate to what type of musical ambiance a music composer could use these sounds, from a calm, neutral passage expressing little emotion to a high-energy musical ambiance composed to evoke tension, excitation or fear.  
For each participant, ratings of negativity were averaged over the 12 matched transformed and non-transformed sounds in each category, and statistical effects were inferred from the sample of participants (N=21) using a rmANOVA with sound category and manipulation (original/ANGUS) as within-item factors. 

\subsection*{Results}
\begin{figure}[ht]
\centering
 \includegraphics[width=\linewidth]{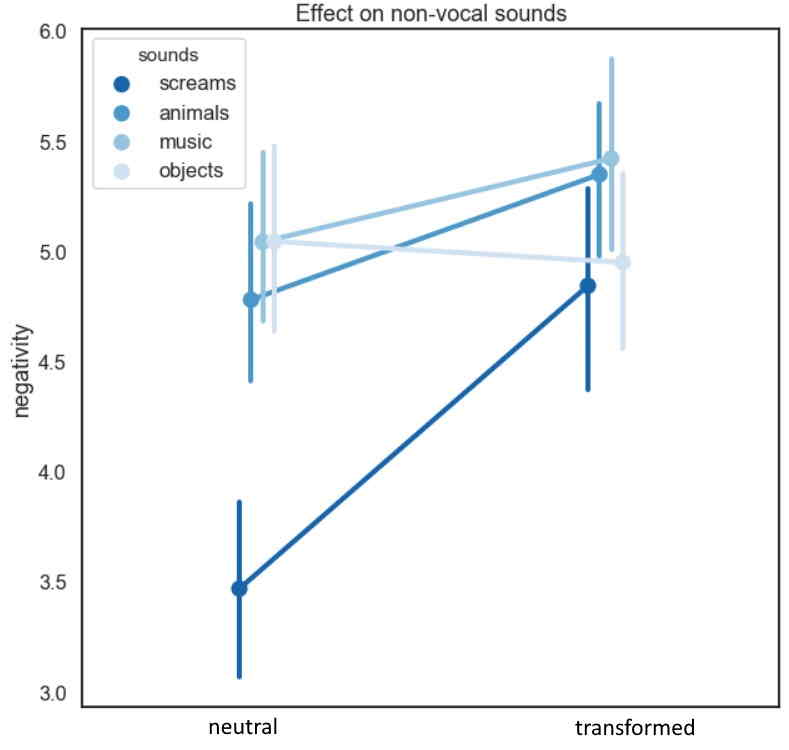}
 \caption{{\bf Effect of ANGUS on the perceived negativity of vocal and non-vocal sounds}. ANGUS had a smaller but similarly negative impact on animal vocalizations and musical instrument sounds than on human vocalizations, but no consistent effect on object sounds. Error bars: 95\% CI on the mean.}
 \label{nonvoc}
\end{figure}

Irrespective of ANGUS, there was a main effect of sound category on perceived negativity (F(3,60)= 10.97, p=7.67e-06), in which human vocalizations were generally evaluated as less negative than other sound categories. More importantly, there was a main effect of the ANGUS manipulation on perceived negativity across sound categories (F(1,20)=65.28, p=9.9e-08) in which, consistent with Study 2, ANGUS increased negativity ratings by $\Delta$=+0.55 [0.33,0.77] scale point. In addition, sound category significantly interacted with the ANGUS effect (F(3,60)=31.48, p=2.36e-12). ANGUS had a more negative effect on human vocalizations ( $\Delta$=+1.37 [1.16,1.59]) than on animal ($\Delta$=+0.57 [0.35 - 0.78]) and musical sounds ($\Delta$=+0.38 [0.16,0.60]). In contrast, ANGUS had no consistent effect on object sounds ( $\Delta$=-0.09, [-0.31,0.12], see Figure \ref{nonvoc}). 

\section*{Discussion}
In this work, we present a real-time acoustic transformation algorithm, ANGUS, which uses amplitude modulation and time-domain filtering to simulate vocal arousal without attempting to directly control jitter and shimmer at the glottal source level. In a series of 4 studies, we show that (1) ANGUS allows parametric, albeit indirect, control over a vocal recording's shimmer as well as its measured jitter, that (2) ANGUS increases perceived emotional negativity of the transformed sounds to a comparable level as a non-real-time analysis/resynthesis algorithm from the state-of-the-art, that (3) listeners cannot distinguish transformed and non-transformed sounds above chance level and (4) that ANGUS has a similar emotional effect on animal vocalizations and musical instrument sounds than on human vocalizations.   

The fact that ANGUS modulates classical measures of jitter and shimmer (Study 1) confirms that subharmonics as created by ANGUS are related to irregularities of the period and amplitude of glottal pulses. It is important to note that the variations possible with ANGUS, and those with direct pulse modeling do not entirely overlap: for instance, frequency modulation of the pulses in fact generates an infinite series of sub-harmonics, rather than a small number as done here, and with more complex amplitude relations than what we model here. In addition, there is more to vocal arousal than roughness, e.g. broadband noise \cite{BLUM12}, pitch jumps \cite{ANIK19} and modification of the spectral slope due to vocal effort \cite{PER16}, which ANGUS does not attempt to simulate. Yet, results from Study 2 suggest that the timbral subspace of vocal arousal explored by ANGUS is sufficient to evoke negative appraisals, and to a degree that is not strikingly less important than complete jitter and shimmer modeling, at least in our dataset. 

Results from Study 2 established that ANGUS and, incidentally, direct jitter and shimmer modeling with the CONTROL algorithm, resulted in a consistent, but relatively small, $\Delta$=+0.5 point increase of perceived negativity on a 10-point scale, compared to non-manipulated sounds. This may be due to the fact that ANGUS only controls vocal timbre (and, even more precisely, only one component of all vocal timbral phenomena associated with vocal arousal, namely roughness), and that there are others, possibly more prominent, cues of emotional negativity that would drive participant judgement in a more ecological situation. For instance, mean pitch alone had here a $\Delta$=+1.0 effect on negativity. Similarly, loudness, which is normalized in the present study, would also contribute to perceived negativity and arousal \cite{BANSE96}. For maximum effect, all of these cues should be combined. 

While results from Study 3 suggest that voices manipulated with ANGUS cannot be reliably discriminated from non-transformed voices in terms of naturalness, it was surprising that voices manipulated with the CONTROL algorithm were judged consistently \emph{more} natural than non-transformed recordings. It may be that including both ANGUS and CONTROL recordings in the same test biased participants to identify only one type of cue for artificiality (e.g. temporal modulation), and to attribute to the 'untransformed' category anything else that did not include these cues (e.g. sounds transformed with the control algorithm). It is also possible that, by using short 600ms extracts, we created a situation in which greatly unstable vocalizations (such as those included in the original stimuli) were difficult to resolve as coherent signals \cite{BON16}, with the result of favoring transformed stimuli that displayed a less complex set of cues within the same time. For a more complete evaluation of how ANGUS affects naturalness, it may be necessary to replicate the study with longer extracts and only one type of manipulated sounds.  

Finally, the results of Study 4 confirm a growing line of research showing that emotional auditory cues such as vocal arousal \cite{BLUM12,ARNA15} but also pitch or spectral content \cite{JUSLIN03,ILIE06,MA15}, are largely transferable across stimulus domains. While the fact that subharmonics result in increased negativity in both human and animal vocalizations is biologically founded, because of cross-species similarities in the acoustic properties of the vocal apparatus and its neural control \cite{FITCH02}, it is more remarkable that cues of vocal arousal should modulate non-vocal musical sounds to the same effect. This is consistent with recent theoretical views according to which the expression of emotions by music, a culturally-evolved phenomenon, exploits biologically-evolved perceptual mechanisms designed to process communicative information in voices and gestures \cite{JUSL08,BRYAN13,SIEV19}. For instance, joyful music is often associated with fast pace and animated pitch contours (as is happy speech), melancholic music with slower and flatter melodic lines and dark timbres (as is sad speech) \cite{JUSLIN03,ILIE06} and, as seen here, exciting music is associated with high levels of roughness, as are angry shouts \cite{OLL19,TRE20}.

In previous research, alarm sounds were made more frightening by adding temporal modulations \cite{ARNA15}. That the same effect is not seen here with alarms and machine sounds requires further investigation: it is possible, e.g., that some of these stimuli did not have clear enough $f_0$ to allow manipulation with ANGUS, or that the transformation interacted with the semantics of the varied sound sources included in this category, making only some of the sounds more negative. It is in fact likely that such context effects affect appraisals of roughness for all types of sounds \cite{BEL15}, allowing e.g. a rough shout to be perceived as a positive bout of laughter \cite{ANIK19}, or rough music as pleasingly empowering \cite{OLL19}. Tools like ANGUS, which allow to control cues of vocal arousal at the stimulus level, contribute to make possible more research on how these cues are cognitively appraised in ecological situations \cite{LIU20}.

On the whole, the present set of results establish that ANGUS is an effective and computationally-efficient method to control vocal arousal in a variety of sounds, for all situations where the focus is less on the precise acoustic control of e.g. pulse characteristics than on the real-time, emotional effect on listeners.   

\section*{Author contributions}
Author ML designed the ANGUS software, and LA designed the CONTROL software. ML, LV and JJA designed the experimental studies. LV and LS collected the data. JJA analysed the data and wrote the manuscript, with contributions from ML and LA. 

\section*{Acknowledgments}
The authors thank Axel R\"obel and Marta Gentilucci (IRCAM) for their comments. Work funded by ERC StG CREAM 335536 and ANR REFLETS (ANR-17-CE19-0020). Experimental data was collected at the Centre Multidisciplinaire des Sciences Comportementales Sorbonne Universit\'es-Institut Europ\'een d’Administration des Affaires (INSEAD). 
 
\bibliography{references.bib}


\nolinenumbers

%
%
%

\end{document}